\documentclass[aps,prc,10pt,showpacs,showkeys,twocolumn,superscriptaddress,groupedaddress]{revtex4-1}
\usepackage[latin1]{inputenc}
\usepackage[english]{babel}
\usepackage{graphicx,psfrag}
\usepackage{amsmath}
\usepackage{amssymb}
\usepackage{amscd}
\usepackage{eucal}
\usepackage{color}
\usepackage{bm}
\usepackage{braket}
\usepackage{dsfont}

\begin{document}
\title{$^7$Be and $^7$Li nuclei within the no-core shell model with continuum}

\author{Matteo Vorabbi}
\email{mvorabbi@triumf.ca}
\affiliation{TRIUMF, 4004 Wesbrook Mall, Vancouver, British Columbia V6T 2A3, Canada}

\author{Petr Navr\'atil}
\email{navratil@triumf.ca}
\affiliation{TRIUMF, 4004 Wesbrook Mall, Vancouver, British Columbia V6T 2A3, Canada}

\author{Sofia Quaglioni}
\affiliation{Lawrence Livermore National Laboratory, P.O. Box 808, L-414, Livermore, California 94551, USA}

\author{Guillaume Hupin}
\affiliation{Institut de Physique Nucl\'eaire, CNRS/IN2P3, Universit\'e Paris-Sud, Universit\'e Paris-Saclay, F-91406, Orsay, France}

\date{\today}

\begin{abstract}

\noindent
{\bf Background:} The production of $^7$Be and $^7$Li nuclei plays an important role in primordial nucleosynthesis, nuclear astrophysics, and fusion energy generation.
The $^3\mathrm{He}(\alpha , \gamma) ^7\mathrm{Be}$ and $^3\mathrm{H}(\alpha , \gamma) ^7\mathrm{Li}$ radiative-capture processes are important to determine the $^7$Li abundance in the early universe and to predict the correct fraction of pp-chain branches resulting in $^7$Be versus $^8$B neutrinos.
The $^6\mathrm{Li}(\mathrm{p},\gamma)^7\mathrm{Be}$ has been investigated recently hinting at a possible cross section  enhacement near the thershold.
The $^6\mathrm{Li}(\mathrm{n},$$^3\mathrm{H}){}^4\mathrm{He}$ process can be utilized for tritium breeding in machines dedicated to fusion energy generation through the deuteron-tritium reaction, and is a neutron cross section standard used in the measurement and evaluation of fission cross sections.

\noindent
{\bf Purpose:} In this work we study the properties of $^7$Be and $^7$Li within the no-core shell model with continuum (NCSMC) method, using chiral nucleon-nucleon interactions
as the only input, and analyze all the binary mass partitions involved in the formation of these systems.

\noindent
{\bf Methods:} The NCSMC is an {\it ab initio} method applicable to light nuclei that provides a unified
description of bound and scattering states and thus is well suited to investigate systems with many resonances and pronounced clustering like $^7$Be and $^7$Li.

\noindent
{\bf Results:} Our calculations reproduce all the experimentally known states of the two systems and provide predictions for several new resonances of both parities.
Some of these new possible resonances are built on the ground states of $^6$Li and $^6$He, and thus represent a robust prediction. We do not find any resonance in the p${+}^6$Li mass partition near the threshold. On the other hand, in the p${+}^6$He mass partition of $^7$Li we observe an $S$-wave resonance near the threshold producing a very pronounced peak in the calculated S factor of the $^6\mathrm{He} (\mathrm{p},\gamma) ^7\mathrm{Li}$ radiative-capture reaction. 

\noindent
{\bf Conclusions:} While we do not find a resonance near the thershold in the p${+}^6$Li channel, in the case of $^6$He${+}$p reaction a resonant $S$-wave state is predicted at a very low energy above the reaction threshold, which could be relevant for astrophysics and its implications should be investigated. We note though that this state lies above the three-body breakup threshold not included in our method and may be influenced by three-body continuum correlations.

\end{abstract}


\maketitle

\section{Introduction}
\label{sec_intro}
The $A{=}7$ systems, in particular $^7$Be and $^7$Li, play an important role in primordial nucleosynthesis, nuclear astrophysics, and fusion energy experiments.

The $^3\mathrm{He}(\alpha , \gamma) ^7\mathrm{Be}$ and $^3\mathrm{H}(\alpha , \gamma) ^7\mathrm{Li}$ radiative-capture processes are crucial for the determination
of the primordial $^7$Li abundance in the early universe~\cite{PhysRevLett.82.4176,PhysRevD.61.123505,PhysRevC.63.054002} and for predicting the correct fraction of
pp-chain branches resulting in $^7$Be versus $^8$B neutrinos~\cite{RevModPhys.70.1265,RevModPhys.83.195}.
Measuring these reactions at the very low solar energies required for astrophysics modeling is extremely challenging due to the suppression of the reaction probability caused by the
Coulomb repulsion between the reactants. Consequently, despite the several experimental measurements~\cite{PhysRevLett.93.262503,PhysRevLett.97.122502,PhysRevC.75.065803,
PhysRevC.76.055801,PhysRevLett.102.232502,PhysRevC.86.032801,BORDEANU20131,PhysRevC.50.2205,PhysRevC.87.065804}, a predictive theoretical description is needed
to reliably guide the extrapolation of higher-energy experimental data down to the desired solar values~\cite{Leva_2016}. A summary of the experimental status of production and
destruction of $^7$Be at the relevant energies for astrophysics can be found in Ref.~\cite{Leva_2016}.

A recent experimental investigation of the $^6\mathrm{Li} (p , \gamma) ^7\mathrm{Be}$ capture reaction at Lanzhou~\cite{HE2013287} hinted at a possible resonant enhancement
of this cross section near the threshold. If real, this enhancement would have consequences for nuclear astrophysics. A new experiment~\cite{BROGGINI201855} is also in progress
at the Laboratory of Underground Nuclear Astrophysics~\cite{luna} (LUNA). A theoretical investigation of $S$-wave resonances in $^7$Be is then called for.

Furthermore, the $^6\mathrm{Li}(n,$$^3\mathrm{H}){}^4\mathrm{He}$ resonant reaction is important for tritium breeding at facilities dedicated to the demonstration of fusion energy
generation with deuterium-tritium fuel such as ITER~\cite{iter}.

Starting from the early 60's several theoretical papers have been devoted to the microscopic description of the $^3\mathrm{He}(\alpha , \gamma) ^7\mathrm{Be}$ and
$^3\mathrm{H}(\alpha , \gamma) ^7\mathrm{Li}$ radiative-capture processes~\cite{PhysRev.131.2582,KAJINO1986559,MERTELMEIER1986387,Csoto2000,
PhysRevLett.106.042502,Gustavino2016,DOHETERALY2016430,bertulani2016,PhysRevC.96.064605,PhysRevC.97.035802,TAKACS201878,Higa2018}.
In particular, in Ref.~\cite{DOHETERALY2016430} these reactions were investigated within the {\it ab initio} No-Core Shell Model with Continuum (NCSMC) using a renormalized chiral nucleon-nucleon interaction. The calculated astrophysical $S$ factors displayed a reasonable agreement with experimental data.

The scope of the present work is to extend this previous investigation of $^7$Be and $^7$Li nuclei published in Ref.~\cite{DOHETERALY2016430} considering all the possible binary mass partitions involved in the formation
of such systems and analyzing the spectra in a wider energy range.

The paper is organized as follows: In Sec.~\ref{sec_formalism} we give the formalism of the NCSMC, while in Sec~\ref{sec_results} we display the results for the $^7$Be and $^7$Li.
We first show the comparison with the experimentally known states and then we will show our predictions for new possible resonant states. Finally, in Sec.~\ref{sec_conclusions} we
draw our conclusions.

\section{Theoretical Framework}
\label{sec_formalism}

Since the aim of this paper is to extend the work of Ref.~\cite{DOHETERALY2016430}, we adopt the same conditions.
The starting point of our approach is the microscopic Hamiltonian
\begin{equation}
\hat{H}=\frac{1}{A}\sum_{i<j=1}^A\frac{(\hat{\vec{p}}_i-\hat{\vec{p}}_j)^2}{2m}+\sum_{i<j=1}^A \hat{V}^{NN}_{ij} \, ,
\label{H}
\end{equation}
which describes nuclei as systems of $A$ non-relativistic point-like nucleons interacting through realistic inter-nucleon interactions.
For consistency with Ref.~\cite{DOHETERALY2016430}, we only use the nucleon-nucleon ($NN$) interaction. Typically, also three-nucleon ($3N$) contributions can be taken into account. In the present work we adopt the $NN$ chiral interaction~\cite{Entem2003} developed by Entem and
Machleidt up to the fourth order (N$^3$LO) in the chiral expansion. In the framework of chiral effective field theory (EFT)~\cite{Weinberg1990,Weinberg1991}, the Lagrangian is
expanded in powers of $(Q/\Lambda_{\chi})^n$, where $Q$ is the external momentum and $\Lambda_{\chi}$ represents the hard scale of the theory and it is chosen of the order
of 1 GeV. Such an expansion allows a systematic improvement of the interaction and provides a hierarchy of the $NN$ and many-nucleon interactions which naturally arise
in a consistent scheme~\cite{OrRa94,VanKolck94,EpNo02,Epelbaum06}.

A faster convergence of the NCSMC calculations is obtained by softening the chiral interaction through the similarity renormalization group (SRG)
technique~\cite{Wegner1994,Bogner2007,PhysRevC.77.064003,Bogner201094,Jurgenson2009}. The general scheme for such a renormalization procedure is to keep two- and
three-body SRG induced terms in all calculations, even in the case when the initial chiral $3N$ force is not included. In the current work, to be consistent with
Ref.~\cite{DOHETERALY2016430} and due to the too high computational effort of dealing with a three-nucleon projectile, we also discard the induced $3N$ terms.

The $NN$ potential is softened via the SRG and we evolved the interaction up to $\lambda_{\mathrm{SRG}} = 2.15$ fm$^{-1}$, where the parameter $\lambda_{\mathrm{SRG}}$
specifies the resolution scale at which the $NN$ potential is evolved. The value of $\lambda_{\mathrm{SRG}}$ adopted in this work is the same as that one used in
Ref.~\cite{DOHETERALY2016430} and it has been chosen as the best value that allows us to reproduce the experimental separation energies of $^7$Be and $^7$Li.

The NCSMC calculation of the scattering observables requires in input the eigenstates of the two colliding nuclei and the eigenstate of the compound system created during
the interaction process. These eigenstates are calculated with the no-core shell model (NCSM) ~\cite{PhysRevLett.84.5728,PhysRevC.62.054311,BARRETT2013131}, an
{\it ab initio} method where all nucleons are considered as active degrees of freedom and the many-body wave function is expanded over a complete set of antisymmetric
$A$-nucleon harmonic-oscillator (HO) basis states containing up to  $N_{\rm max}$ HO excitations above the lowest Pauli-principle-allowed configuration:
\begin{equation}\label{NCSM_wav}
 \ket{\Psi^{J^\pi T}_A} = \sum_{N=0}^{N_{\rm max}}\sum_i c_{Ni}^{J^\pi T}\ket{ANiJ^\pi T} \, .
\end{equation}
Here, $N$ denotes the total number of HO excitations of all nucleons above the minimum configuration,  $J^\pi T$ are the total angular momentum, parity and isospin,
and $i$ additional quantum numbers. The sum over $N$ is restricted by parity to either an even or odd sequence. The basis is further characterized by the frequency $\Omega$
of the HO well. Square-integrable energy eigenstates expanded over the $N_{\rm max}\hbar\Omega$ basis, $\ket{ANiJ^\pi T}$, are obtained by diagonalizing the intrinsic
Hamiltonian of Eq.(\ref{H}). In this work the value of the HO frequency has been chosen to be $\hbar \Omega = 20$ MeV, again consistently with Ref.~\cite{DOHETERALY2016430}.

The NCSMC wave function is then represented as the generalized cluster expansion. For $^7$Be we have
\begin{align}
\ket{\Psi^{J^\pi T}_{A\texttt{=}7,\frac{1}{2}}} = &  \sum_\lambda c^{J^\pi T}_\lambda \ket{^{7} {\rm Be} \, \lambda J^\pi T} \nonumber \\
& +\sum_{\nu}\!\! \int \!\! dr \, r^2 
                 \frac{\gamma^{J^\pi T}_{\nu}(r)}{r}
                 {\mathcal{A}}_\nu \ket{\Phi^{J^\pi T}_{\nu r, \frac{1}{2}}} \,,\label{ncsmc_wf_Be}
\end{align}
while for $^7$Li we have
\begin{align}
\ket{\Psi^{J^\pi T}_{A\texttt{=}7, -\frac{1}{2}}} = &  \sum_\lambda c^{J^\pi T}_\lambda \ket{^{7} {\rm Li} \, \lambda J^\pi T} \nonumber \\
& +\sum_{\nu}\!\! \int \!\! dr \, r^2 
                 \frac{\gamma^{J^\pi T}_{\nu}(r)}{r}
                 {\mathcal{A}}_\nu \ket{\Phi^{J^\pi T}_{\nu r, -\frac{1}{2}}} \,,\label{ncsmc_wf_Li}
\end{align}
The first term of Eqs.(\ref{ncsmc_wf_Be}) and (\ref{ncsmc_wf_Li}) consists of an expansion over NCSM eigenstates of the aggregate system ($^{7}$Be and $^7$Li) indexed
by $\lambda$.
These states are well suited to explain the localized correlations of the two 7-body systems, but are inadequate to describe clustering and scattering properties that
are better addressed by the second term corresponding to an expansion over the antisymmetrized channel states in the spirit of the resonating group
method~\cite{wildermuth1977unified,TANG1978167,FLIESSBACH198284,langanke1986,PhysRevC.77.044002}. For $^7$Be we have
\begin{align}
\ket{\Phi^{J^\pi T}_{\nu r, \frac{1}{2}}} = &\Big[ \big( \ket{^{4} {\rm He} \, \lambda_4 J_4^{\pi_4}T_4} \ket{^{3} {\rm He} \, \lambda_3 J_3^{\pi_3}T_3} \big)^{(sT)} \nonumber \\
&\times \, Y_\ell(\hat{r}_{4,3}) \Big]^{(J^{\pi}T)}_{\frac{1}{2}} \; \frac{\delta(r{-}r_{4,3})}{rr_{4,3}} \, , \label{eq_Be_rgm_state1} \\
\ket{\Phi^{J^\pi T}_{\nu r, \frac{1}{2}}} = &\Big[ \big( \ket{^{6} {\rm Li} \, \lambda_6 J_6^{\pi_6}T_6} \ket{p \, \tfrac12^{\texttt{+}}\tfrac12} \big)^{(sT)}
Y_\ell(\hat{r}_{6,1}) \Big]^{(J^{\pi}T)}_{\frac{1}{2}} \nonumber\\ 
&\times\,\frac{\delta(r{-}r_{6,1})}{rr_{6,1}} \, ,
\label{eq_Be_rgm_state2}
\end{align}
while for $^7$Li we have
\begin{align}
\ket{\Phi^{J^\pi T}_{\nu r, -\frac{1}{2}}} = &\Big[ \big( \ket{^{4} {\rm He} \, \lambda_4 J_4^{\pi_4}T_4} \ket{^{3} {\rm H} \, \lambda_3 J_3^{\pi_3}T_3} \big)^{(sT)} \nonumber \\
&\times \, Y_\ell(\hat{r}_{4,3}) \Big]^{(J^{\pi}T)}_{-\frac{1}{2}} \; \frac{\delta(r{-}r_{4,3})}{rr_{4,3}} \, , \\
\ket{\Phi^{J^\pi T}_{\nu r, -\frac{1}{2}}} = &\Big[ \big( \ket{^{6} {\rm Li} \, \lambda_6 J_6^{\pi_6}T_6} \ket{n \, \tfrac12^{\texttt{+}}\tfrac12} \big)^{(sT)}
Y_\ell(\hat{r}_{6,1}) \Big]^{(J^{\pi}T)}_{-\frac{1}{2}} \nonumber\\ 
&\times\,\frac{\delta(r{-}r_{6,1})}{rr_{6,1}} \, , \\
\ket{\Phi^{J^\pi T}_{\nu r, -\frac{1}{2}}} = &\Big[ \big( \ket{^{6} {\rm He} \, \lambda_6 J_6^{\pi_6}T_6} \ket{p \, \tfrac12^{\texttt{+}}\tfrac12} \big)^{(sT)}
Y_\ell(\hat{r}_{6,1}) \Big]^{(J^{\pi}T)}_{-\frac{1}{2}} \nonumber\\ 
&\times\,\frac{\delta(r{-}r_{6,1})}{rr_{6,1}} \, .
\label{eq_Li_rgm_state}
\end{align}
The $\nu$ index represents all the quantum numbers on the right-hand side not appearing on the left-hand side and the subscript $\pm\frac{1}{2}$ is the isospin projection, i.e., $(Z-N)/2$. All these terms describe the relative motion of the two colliding nuclei involved in the formation of the composite system during the scattering process. Here, the coordinate
$\vec{r}_{4,3}$ in Eq.(\ref{eq_Be_rgm_state1}) is the separation distance between the center-of-mass of $^4$He and $^3$He, while $\vec{r}_{6,1}$ in Eq.(\ref{eq_Be_rgm_state2})
represents the separation distance between $^6$Li and the proton. The same meaning holds for the separation distances in the channel states of $^7$Li.

The discrete expansion coefficients  $c_{\lambda}^{J^{\pi}T}$ and the continuous relative-motion amplitudes $\gamma_{\nu}^{J^{\pi}T} (r)$ are the solution of the generalized
eigenvalue problem derived by representing the Schr\"{o}dinger equation in the model space of the expansions (\ref{ncsmc_wf_Be}) and (\ref{ncsmc_wf_Li})~\cite{physcripnavratil}.
The resulting NCSMC equations are solved by the coupled-channel R-matrix method on a Lagrange mesh~\cite{Descouvemont2010,Hesse1998,Hesse2002}.
We emphasize that the sums over the index $\nu$ in Eqs.(\ref{ncsmc_wf_Be}) and (\ref{ncsmc_wf_Li}) include all the mass partitions involved in the formation of the compound
systems $^7$Be and $^7$Li. The NCSMC calculation of these systems with the coupling between the all binary mass partitions is however beyond our present capabilities
due to the challenge of dealing with a three-body projectile involved in the $^3$He + $^4$He and $^3$H + $^4$He reactions. In the present work we will thus consider the different
partitions separately. Applications of the NCSMC with a two-body projectile and with the coupling between different mass partitions can be found
in Refs.~\cite{PhysRevC.93.054606,Hupin2019}.

\section{Results}
\label{sec_results}

The NCSMC calculations require in input the NCSM eigenstates of the colliding nuclei and of the composite system. In this work we used eight lowest negative-parity and six lowest positive-parity
NCSM eigenstates for $^7$Be and $^7$Li with total angular momentum  $J \in \{ 1/2 , 3/2 , 5/2 , 7/2 \}$ and isospin $T = 1/2$. Concerning the reactants, we used the ground
state for $^4$He $[(J^{\pi} T) = (0^+ 0) ]$ and $^3$He ($^3$H) $[(J^{\pi} T) = 1/2^+ 1/2]$, while we used four states for
$^6$Li $[(J^{\pi} T) = (1^+ 0) , (3^+ 0) , (0^+ 1) , (2^+ 1)]$ and two states for $^6$He $[(J^{\pi} T) = (0^+ 1) , (2^+ 1)]$. All the calculations were performed using the SRG evolved chiral N$^3$LO $NN$ interaction~\cite{Entem2003} with $\lambda_{\mathrm{SRG}} = 2.15$ fm$^{-1}$ and with $\hbar \Omega = 20$ MeV. In the following subsections we present results obtained at $N_{\mathrm{max}} = 11$, the largest space we could reach for technical reasons. However, we performed calculations also at smaller $N_{\mathrm{max}}$ spaces to check convergence.

Every single mass partition was studied separately, neglecting the coupling with the other mass partitions. Depending on the reaction under consideration,
the states of the composite system that are below the reaction threshold come out from our calculations as bound states. In the following we will thus show the results for
$^7$Be and $^7$Li in two separate subsections, presenting first the values of the ground state ($3/2_1^-$) and the first excited state ($1/2_1^-$) of both systems obtained from the
calculation performed with the different mass partitions. We will then show the results for the resonant states and the comparison with the experimentally known ones, and we will finally show our predictions for new possible resonances.

\subsection{The $^7$Be system}

\begin{table}[t]
\begin{center}
\begin{ruledtabular}
\begin{tabular}{c|cc}
$^3$He + $^4$He $\; \leftrightarrow \;$  $^7$Be \hspace{0.1cm} & $J^{\pi} = 3/2^-$ & $J^{\pi} = 1/2^-$ \hspace{0.5cm} \\
\hline
$E_B$ \hspace{0.15cm} [MeV] & -1.52 & -1.26 \hspace{0.5cm} \\
Exp. [MeV] & -1.587 & -1.157 \hspace{0.5cm} \\
\hline
$E$ \hspace{0.35cm} [MeV] & -36.98 & -36.71 \hspace{0.5cm} \\
Exp. [MeV] & -37.60 & -37.17 \hspace{0.5cm} \\
\hline
\hline
\hspace{0.23cm} $^6$Li + p $\; \leftrightarrow \;$  $^7$Be \hspace{0.23cm} & $J^{\pi} = 3/2^-$ & $J^{\pi} = 1/2^-$ \hspace{0.5cm} \\
\hline
$E_B$ \hspace{0.15cm} [MeV] & -5.73 & -5.39 \hspace{0.5cm} \\
Exp. [MeV] & -5.606 & -5.177 \hspace{0.5cm} \\
\hline
$E$ \hspace{0.35cm} [MeV] & -36.47 & -36.13 \hspace{0.5cm} \\
Exp. [MeV] & -37.60 & -37.17 \hspace{0.5cm}
\end{tabular}
\end{ruledtabular}
\caption{Comparison between the NCSMC and the experimental relative ($E_B$) and total ($E$) energies of the $J^{\pi} = 3/2_1^-$ and $1/2_1^-$ bound states of $^7$Be produced with
the $^3$He + $^4$He and $^6$Li + p reactions. Here, the $E_B$ energy represents the energy of the state with respect the threshold of the reaction.
All calculations were performed at $N_{\mathrm{max}} = 11$ in the HO expansion and using the SRG-evolved NN N$^3$LO chiral interaction~\cite{Entem2003} at the resolution
scale of $\lambda_{\mathrm{SRG}} = 2.15$ fm$^{-1}$.}  
\label{be_bound}
\end{center}
\end{table}

\begin{table}[t]
\begin{center}
\begin{ruledtabular}
\begin{tabular}{ccccc}
$^3$He + $^4$He & NCSM & NCSMC & Exp. & Refs. \\
\hline
$r_{\mathrm{ch}}$ [fm] & 2.38 & 2.62 & 2.647(17) & \cite{PhysRevLett.102.062503} \\
$r_{\mathrm{m}}$ [fm] & 2.15 & 2.41 & 2.42(4) & \cite{DOBROVOLSKY201940} \\
$Q$ [e fm$^2$] & -4.57 & -6.14 & - & \\
$\mu$ [$\mu_N$] & -1.14 & -1.16 & -1.3995(5) & \cite{PhysRevLett.102.062503} \\
B(E2) [e$^2$ fm$^4$]  & 10.90 & 20.02 & 26(6)(3) & \cite{notredame}\\
B(M1) [$\mu_N^2$]  & 1.55 & 1.49 & - & \\
$(C_{p3/2})^2$ [fm$^{-1}$] & - & 15.78 & $23.3^{+1.0}_{-2.3}$ & \cite{PhysRevC.85.045807} \\ 
$(C_{p1/2})^2$ [fm$^{-1}$] & - & 13.22 & $15.9^{+0.6}_{-1.5}$ & \cite{PhysRevC.85.045807} 
\end{tabular}
\end{ruledtabular}
\caption{Properties of the ground state of $^7$Be computed with the NCSM and NCSMC approaches using the $^3$He + $^4$He mass partition and
compared with the experimental data. The reduced transition probabilities are from the ground state $3/2_1^-$ to the first excited state $1/2_1^-$. The ANCs $C_{lj}$ are shown for both the ground state and the $1/2^-_1$ state.}  
\label{be_moments}
\end{center}
\end{table}

\begin{figure}[tb]
\begin{center}
\includegraphics[scale=0.33]{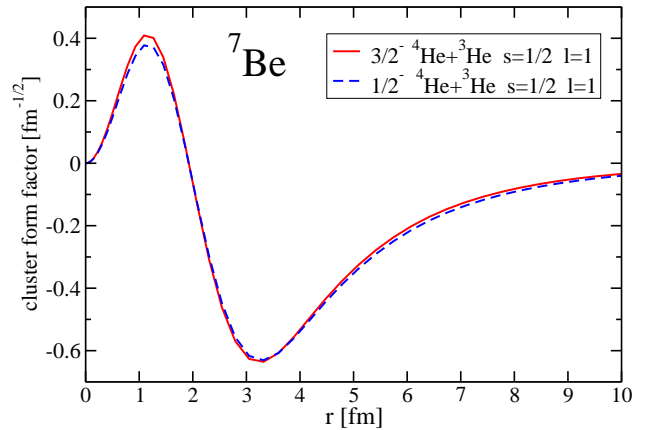}
\caption{\label{be_clusterff}  (Color online) NCSMC cluster form factor of the $^7$Be ground state (solid line) and the $1/2^-_1$ state (dashed line) calculated for the $^4$He+$^3$He mass partition.}
\end{center}
\end{figure}

\begin{figure*}[t]
\begin{center}
\includegraphics[scale=0.65]{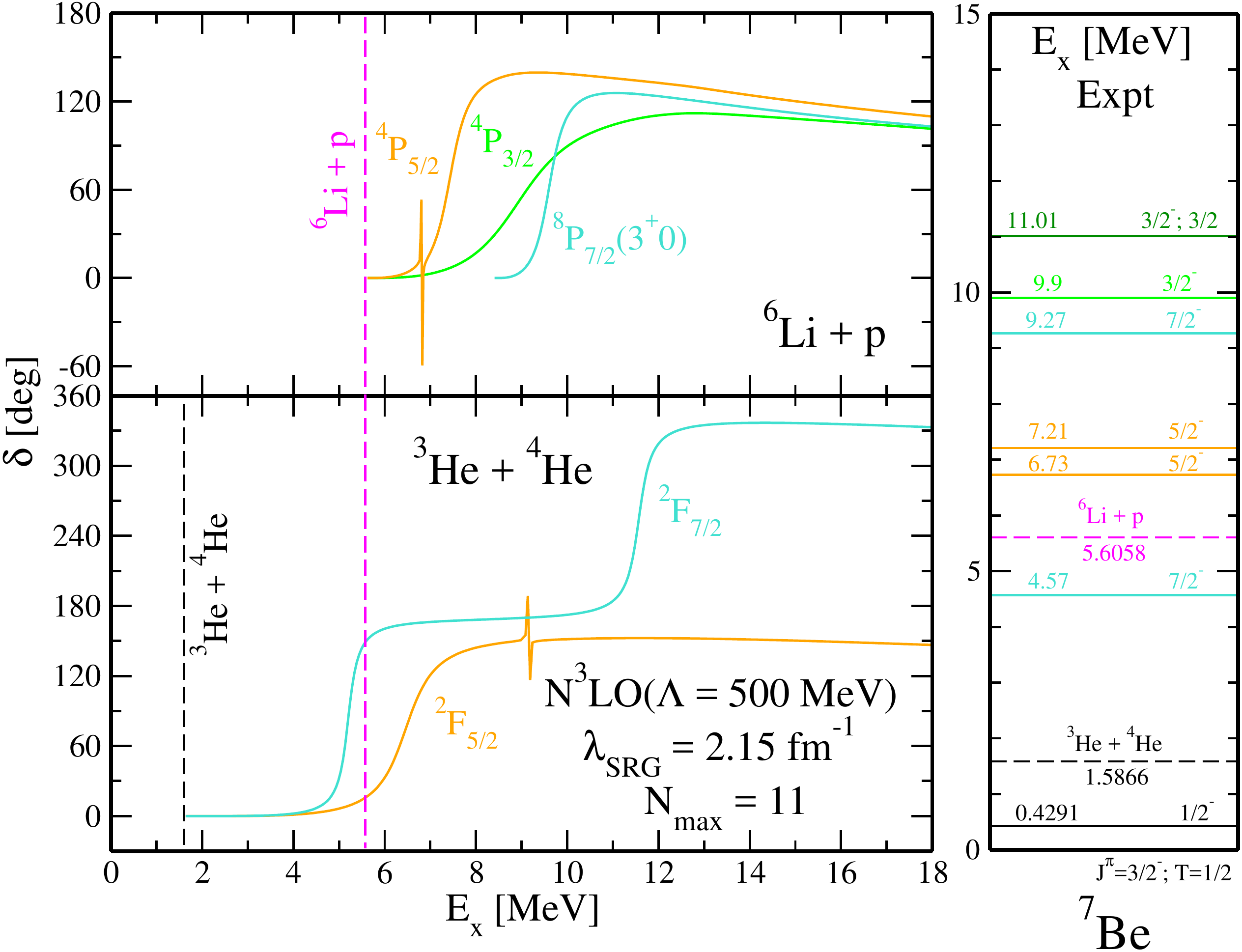}
\caption{\label{spectrum_be}  (Color online) Comparison between the theoretical and the experimental energy spectrum of $^7$Be. The two panels on the left show the
theoretical phase shifts of $^3$He + $^4$He and $^6$Li + p scattering computed with the {\it ab initio} NCSMC method, while the column on the right displays the
experimental energy spectrum. The solid lines represents the energy states while the dashed lines show the thresholds of the two processes.
The theoretical results have been adjusted to the experimental excitation energy. The calculated thresholds are shown in Table~\ref{be_bound}.
In the partial wave labels we also provide information on the target state for phase shifts not built on the ground state.
Calculations were performed at $N_{\mathrm{max}} = 11$ in the HO expansion and
using the SRG-evolved NN N$^3$LO chiral interaction~\cite{Entem2003} at the resolution scale of $\lambda_{\mathrm{SRG}} = 2.15$ fm$^{-1}$.}
\end{center}
\end{figure*}

\begin{figure}[t]
\begin{center}
\includegraphics[scale=0.4]{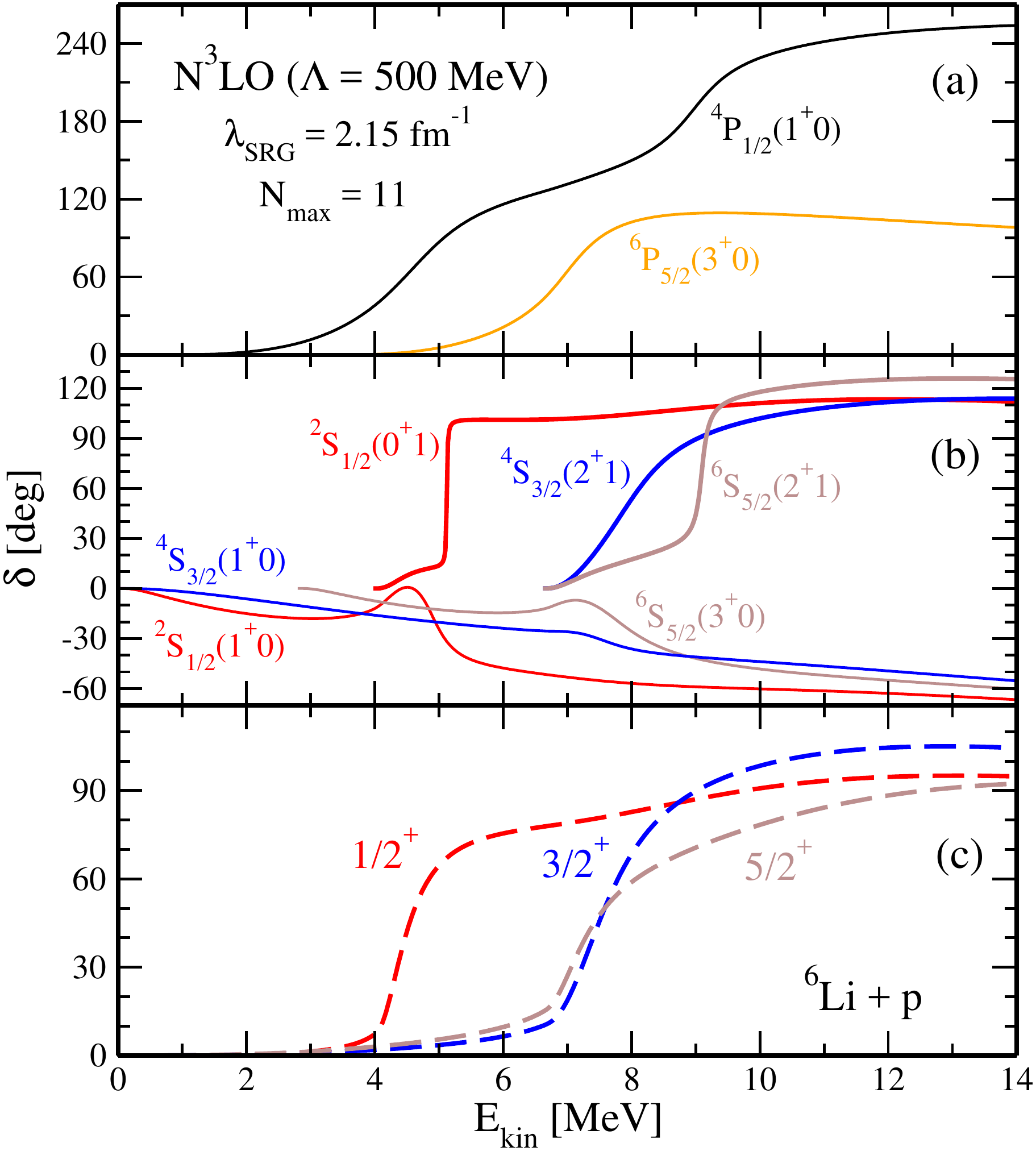}
\caption{\label{be_new_states}  (Color online) Predictions of new negative- and positive-parity resonant states in the spectrum of $^7$Be obtained from the NCSMC calculation of the $^6$Li + p scattering. The negative-parity (positive-parity) phase shifts are displayed in panels (a) and (b), respectively. In panel (c) we present
eigenphase shifts corresponding to the positive-parity resonant states shown in panel (b). All shown partial waves have isospin $T=1/2$.
In the phase shift partial wave labels we also provide information on $^6$Li states.
The results are displayed as functions of the kinetic energy in the center-of-mass. Calculations were performed at $N_{\mathrm{max}} = 11$ in the HO expansion and using the SRG-evolved NN N$^3$LO chiral interaction~\cite{Entem2003} at the resolution scale of $\lambda_{\mathrm{SRG}} = 2.15$ fm$^{-1}$.}
\end{center}
\end{figure}

\begin{table}[t]
\begin{center}
\begin{ruledtabular}
\begin{tabular}{cccccc}
$^3$He + $^4$He & NCSMC & & Exp. & & \\
\hline
$J^{\pi}$ & $E_r$ & $\Gamma$ & $E_r$ & $E_x$ & $\Gamma$ \\
$7/2_1^-$ & 3.61 & 0.33 & 2.98 & 4.57 & 0.175 \\
$5/2_1^-$ & 4.87 & 1.00 & 5.14 & 6.73 & 1.2  \\
$7/2_2^-$ & 9.98 & 0.40 & 7.68 & 9.27 & - \\
\hline
\hline
$^6$Li + p & NCSMC & & Exp. & & \\
\hline
$J^{\pi}$ & $E_r$ & $\Gamma$ & $E_r$ & $E_x$ & $\Gamma$ \\
$5/2_2^-$ & 1.83 & 0.66 & 1.60 & 7.21 & 0.40 \\
$7/2_2^-$ & 3.99 & 0.60 & 3.66  & 9.27 & - \\
$3/2_2^-$ & 3.24 & 2.19 & 4.29  & 9.9 & 1.8 \\
\end{tabular}
\end{ruledtabular}
\caption{Energies ($E_r$) and widths ($\Gamma$) in MeV of the resonant states of $^7$Be computed with the NCSMC and compared with the existing experimental
data~\cite{TILLEY20023}. Only the resonances with $T=1/2$ are considered and the resonance energies are given with respect the threshold of the corresponding
reaction.}  
\label{be_resonances}
\end{center}
\end{table}

\begin{table}[t]
\begin{center}
\begin{ruledtabular}
\begin{tabular}{ccc}
\hspace{0.6cm} $^6$Li + p & NCSMC & \hspace{0.6cm} \\
\hline
\hspace{0.6cm} $J^{\pi}$ & $E_r$ & $\Gamma$ \hspace{0.6cm} \\
\hspace{0.6cm} $1/2^-$ & 4.11 & 2.15 \hspace{0.6cm} \\
\hspace{0.6cm} $5/2^-$ & 6.79 & 4.47 \hspace{0.6cm} \\
\hspace{0.6cm} $1/2^+$ & 4.22 & 0.96 \hspace{0.6cm} \\
\hspace{0.6cm} $3/2^+$ & 7.26 & 1.69 \hspace{0.6cm} \\
\end{tabular}
\end{ruledtabular}
\caption{Energies ($E_r$) and widths ($\Gamma$) in MeV of some of the new predicted resonant states of $^7$Be computed with the NCSMC. Only the resonances with $T=1/2$ are
considered and the resonance energies are given with respect the threshold of the $^6$Li + p mass partition.}  
\label{be_new_resonances}
\end{center}
\end{table}

We studied the $^7$Be nucleus within the NCSMC using the $^3$He + $^4$He and $^6$Li + p reactions, which represent the only two binary mass partitions in this system. All the other mass partitions involve at least three reactants and are not considered in the present work.

The relative ($E_B$) and total ($E$) energies of the $J^{\pi} = 3/2_1^-$ and $1/2_1^-$ states of $^7$Be computed with the NCSMC are displayed in Tab.~\ref{be_bound} and are
compared with the corresponding experimental values. Here, the $E_B$ energies represent the energy of the state under consideration
with respect the threshold of the considered reaction. We can see that in both cases, not only the $E_B$ energies, but also the total ones are in a very good agreement with the
experimental values. This is particularly true for the $^3$He + $^4$He reaction, where the difference between the theoretical and the experimental total energies is less than 1 MeV.

In Tab.~\ref{be_moments} we also report the properties of the ground sate of $^7$Be obtained from the study of the $^3$He + $^4$He reaction with the NCSM and NCSMC,
and we compare our results with the existing experimental data. Besides the values of the charge radius ($r_{\mathrm{ch}}$), quadrupole moment ($Q$), and
magnetic dipole moment ($\mu$), already reported in Ref.~\cite{DOHETERALY2016430}, Tab.~\ref{be_moments} is now updated with the theoretical predictions of the matter radius and the reduced transition probabilities B(M1; $3/2_1^- \rightarrow 1/2_1^-$) and B(E2; $3/2_1^- \rightarrow 1/2_1^-$). The experimental measurement of the B(E2) value has been recently performed at the University of Notre Dame~\cite{notredame}. Our NCSMC result is consistent with this new measurement. Also recently, a new matter radius measurement was published in Ref.~\cite{DOBROVOLSKY201940}. Our calculation is in an excellent agreement with the reported value. In addition, we also present the asymptotic normalization coefficients (ANCs) for both the ground state and the $1/2^-_1$ bound state. The cluster form factors for the two states defined by $r \bra{\Phi^{J^\pi T}_{\nu r, \frac{1}{2}}} {\mathcal{A}}_\nu\ket{\Psi^{J^\pi T}_{A\texttt{=}7,\frac{1}{2}}}$ with the ket and bra from Eqs.~(\ref{ncsmc_wf_Be}) and (\ref{eq_Be_rgm_state1}), respectively, are shown in Fig.~\ref{be_clusterff}. We can see that the NCSMC wave functions extend beyond 10 fm. 

In Fig.~\ref{spectrum_be} we show the phase shifts computed with the NCSMC and compared to the experimental spectrum of $^7$Be. On the left-hand side of the figure
we show the results for the two mass partitions. The lower panel on the left contains the phase shifts for the $^3$He + $^4$He reaction,
while the upper panel contains the results obtained from the $^6$Li + p calculation. The phase shifts are displayed as functions of the experimental excitation energy and the
experimental values of the thresholds are displayed with the two vertical dashed lines. The comparison between the theoretical and the experimental resonance energies
are shown in Tab.~\ref{be_resonances}.
The column on the right displays the experimental excitation energy spectrum. The solid lines are the known resonant states with their relative energies on the left and
their $J^{\pi}$ quantum numbers on the right, while the dashed lines show the thresholds of the two processes, which match with the dashed lines in the left panels.

From Fig.~\ref{spectrum_be} we can clearly see that our results are in a very good agreement with the experimental energy spectrum and all the states are reproduced
in the correct order. In the spectrum of the $^3$He + $^4$He reaction we have four phase shifts corresponding to the $7/2_1^- , 5/2_1^- , 5/2_2^- , 7/2_2^-$
states, while for $^6$Li + p we have the $5/2_1^- , 5/2_2^- , 7/2_2^- , 3/2_2^-$ states.
The $7/2_1^-$ state is below the $^6$Li + p threshold and thus it is only showed in the spectrum of the $^3$He + $^4$He process. Of course this state comes out from the
$^6$Li + p NCSMC calculation as a bound state, exactly as it happens for the $3/2_1^-$ and the $1/2_1^-$ ones. The $3/2_3^-$ state with $T=3/2$ at $11.01$ MeV is not shown
here and it will be discussed in the mirror $^7$Li system. Finally, the sharp peak in the $^2\mathrm{F}_{5/2}$ phase shift of the $^3$He + $^4$He spectrum and
the sharp peak in the $^2\mathrm{P}_{5/2}$ phase shift of the $^6$Li + p spectrum deserve a comment. Experimentally, the cross section of the $^3$He + $^4$He process only
shows a peak in correspondence of the energy of the $5/2_1^-$ state, while for $^6$Li + p a peak is only found at the energy of the $5/2_2^-$ state.
Thus, the contribution of the $^3$He + $^4$He to the $5/2_2^-$ state is basically negligible, exactly as it happens for the $5/2_1^-$ state in the $^6$Li + p spectrum.
Even without the coupling between the two mass partitions, our results correctly reproduce the contributions to the two $5/2^-$ states in both spectra.

In Tab.~\ref{be_resonances} we display the numerical values of the energies and widths of the known states computed with the NCSMC and we compare our results with the
experimental values, where we also show the experimental excitation energy, for a better comparison with Fig.~\ref{spectrum_be}. Both energies and widths are nicely reproduced
even though the agreement with the experimental data is not perfect. For the $^3$He + $^4$He mass partition we did not include in the table the $5/2_2^-$ state, while
for $^6$Li + p we left out the $5/2_1^-$ state. These two states are indeed too narrow and the widths are negligible in those respective mass partitions.

In Fig.~\ref{be_new_states} we show our predictions for new possible $T=1/2$ resonances that are experimentally unknown. As seen in the experimental spectrum in Fig.~\ref{spectrum_be}, all known states are of negative parity. The interesting result is that we found new possible resonances of both parities.
In panel (a) we present the $^4\mathrm{P}_{1/2}$ and $^6\mathrm{P}_{5/2}$ phase shifts built on the $^6$Li ground state and the $3^+ 0$ excited state, respectively. We note that the first $^4\mathrm{P}_{1/2}$ resonance has been observed experimentally in $^7$Li, see Sect.~\ref{Li7}. Consequently, we expect that it should be possible to observe it in $^7$Be as well. The $S$-wave phase shifts built on the four $^6$Li states included in our calculations are plotted in panel (b). We observe a resonance behavior in particular in the phase shifts built on the $0^+$ and $2^+$ $T{=}1$ states. The corresponding eigenphase shifts are displayed in the panel (c) and confirm resonances in $1/2^+, 3/2^+$ and $5/2^+$ channels. The resonance energies and widths of some of these predicted resonances are summarized in Tab.~\ref{be_new_resonances}. We note that all these predicted resonances are in three-body continuum not included in our calculations. In particular, we anticipate that the $3/2^+$ and $5/2^+$ will be influenced by the three-body continuum as the $2^+ 1$ states of $^6$Li is rather broad.

\subsubsection*{The $^6\mathrm{Li} (p , \gamma) ^7\mathrm{Be}$ radiative capture reaction}

As discussed in the Introduction, recently the $^6\mathrm{Li} (p , \gamma) ^7\mathrm{Be}$ capture reaction has been investigated at Lanzhou~\cite{HE2013287} and suggested a possible new structure just above the threshold that has not been observed in the previous measurements~\cite{SWITKOWSKI197950,Tingwell1987,CECIL199275,PhysRevC.70.055801}. In our analysis, we also computed the capture reaction. We do not find any resonance near the $^6$Li${+}$p threshold as shown in Figs.~\ref{spectrum_be} and ~\ref{be_new_states}. Consequently, the shape of our calculated $S$ factor reproduce the trend of the earlier experimental data with no evidence of a new structure at very low energy. Recently, the same reaction has been also investigated in Ref.~\cite{Dong_2017,Gnech:2019usu} with similar findings. As in our calculation the coupling between the $^6$Li${+}$p and $^3$He${+}^4$He mass partitions is not included, our calculated $S$ factor overestimates the data. The extension of our formalism to include the coupling is under way. It is clear that it will improve the description of the magnitude of the $S$ factor but it will not change its shape.

\subsubsection*{The $^3\mathrm{He} (\alpha , \gamma) ^7\mathrm{Be}$ radiative capture reaction}

The S factor of the $^3\mathrm{He} (\alpha , \gamma) ^7\mathrm{Be}$ radiative capture calculated with the same Hamiltonian and approach as employed in this paper was presented for kinetic energy range up to 3.8 MeV in the center of mass by black solid line in the top panel of Fig.~5 of Ref.~\cite{DOHETERALY2016430}. Recently,  motivated by the above discussed Lanzhou experiment a new $^3\mathrm{He} (\alpha , \gamma) ^7\mathrm{Be}$ measurement has been performed in the energy range between 4 and 4.5 MeV~\cite{PhysRevC.99.055804}, i.e., in the region just above the threshold of the $^6\mathrm{Li} (p , \gamma) ^7\mathrm{Be}$ capture. This first experiment above 3.1 MeV in the center of mass found a flat structureless S factor. For example, at 4.42 MeV, the $S_{34}=0.55(4)$ keV b has been reported~\cite{PhysRevC.99.055804}. Our calculated NCSMC S factor beyond the range shown in Ref.~\cite{DOHETERALY2016430} is monotonically increasing; at 4.42 MeV, we find $S_{34}=0.48$ keV b, which is slightly below the experiment~\cite{PhysRevC.99.055804}.

\subsection{The $^7$Li system}\label{Li7}

\begin{table}[t]
\begin{center}
\begin{ruledtabular}
\begin{tabular}{c|cc}
$^3$H + $^4$He $\; \leftrightarrow \;$  $^7$Li \hspace{0.1cm} & $J^{\pi} = 3/2^-$ & $J^{\pi} = 1/2^-$ \hspace{0.5cm} \\
\hline
$E_B$ \hspace{0.15cm} [MeV] & -2.43 & -2.15 \hspace{0.5cm} \\
Exp. [MeV] & -2.467 & -1.989 \hspace{0.5cm} \\
\hline
$E$ \hspace{0.35cm} [MeV] & -38.65 & -38.37 \hspace{0.5cm} \\
Exp. [MeV] & -39.25 & -38.77 \hspace{0.5cm} \\
\hline
\hline
\hspace{0.23cm} $^6$Li + n $\; \leftrightarrow \;$  $^7$Li \hspace{0.23cm} & $J^{\pi} = 3/2^-$ & $J^{\pi} = 1/2^-$ \hspace{0.5cm} \\
\hline
$E_B$ \hspace{0.15cm} [MeV] & -7.38 & -7.05 \hspace{0.5cm} \\
Exp. [MeV] & -7.250 & -6.772 \hspace{0.5cm} \\
\hline
$E$ \hspace{0.35cm} [MeV] & -38.13 & -37.79 \hspace{0.5cm} \\
Exp. [MeV] & -39.25 & -38.77 \hspace{0.5cm} \\
\hline
\hline
\hspace{0.23cm} $^6$He + p $\; \leftrightarrow \;$  $^7$Li \hspace{0.23cm} & $J^{\pi} = 3/2^-$ & $J^{\pi} = 1/2^-$ \hspace{0.5cm} \\
\hline
$E_B$ \hspace{0.15cm} [MeV] & -10.39 & -10.06 \hspace{0.5cm} \\
Exp. [MeV] & -9.975 & -9.498 \hspace{0.5cm} \\
\hline
$E$ \hspace{0.35cm} [MeV] & -38.06 & -37.73 \hspace{0.5cm} \\
Exp. [MeV] & -39.25 & -38.77 \hspace{0.5cm}
\end{tabular}
\end{ruledtabular}
\caption{Comparison between the NCSMC and the experimental relative ($E_B$) and total ($E$) energies of the $J^{\pi} = 3/2_1^-$ and $1/2_1^-$ bound states of $^7$Li produced with
the $^3$H + $^4$He, $^6$Li + n, and $^6$He + p reactions. Here, the $E_B$ energy represents the energy of the state with respect the threshold of the reaction.
All calculations were performed at $N_{\mathrm{max}} = 11$ in the HO expansion and using the SRG-evolved NN N$^3$LO chiral interaction~\cite{Entem2003} at the resolution
scale of $\lambda_{\mathrm{SRG}} = 2.15$ fm$^{-1}$.}  
\label{li_bound}
\end{center}
\end{table}

\begin{table}[t]
\begin{center}
\begin{ruledtabular}
\begin{tabular}{ccccc}
$^3$H + $^4$He & NCSM & NCSMC & Exp. & Refs. \\
\hline
$r_{\mathrm{ch}}$ [fm] & 2.21 & 2.42 & 2.39(3) & \cite{DEJAGER1974479} \\
$Q$ [e fm$^2$] & -2.67 & -3.72 & -4.00(3) & \cite{VOELK1991475} \\
$\mu$ [$\mu_N$] & 3.00 & 3.02 & 3.256 & \cite{RAGHAVAN1989189} \\
B(E2) [e$^2$ fm$^4$]  & 3.49 & 7.12 & 8.3(5) & \cite{PhysRevLett.55.480} \\
B(M1) [$\mu_N^2$]  & 2.05 & 2.00 & - & \\
$(C_{p3/2})^2$ [fm$^{-1}$] & - & 12.21 & 12.74(110) & \cite{NuclPhysA.781.247} \\ 
$(C_{p1/2})^2$ [fm$^{-1}$] & - &   9.99 & 9.00(90) & \cite{NuclPhysA.781.247} 
\end{tabular}
\end{ruledtabular}
\caption{Properties of the ground state of $^7$Li computed with the NCSM and NCSMC approaches using the $^3$H + $^4$He mass partition and
compared with the experimental data. The reduced transition probabilities are from the ground state $3/2_1^-$ to the first excited state $1/2_1^-$. 
The ANCs $C_{lj}$ are shown for both the ground state and the $1/2^-_1$ state.}  
\label{li_moments}
\end{center}
\end{table}

\begin{figure*}[t]
\begin{center}
\includegraphics[scale=0.68]{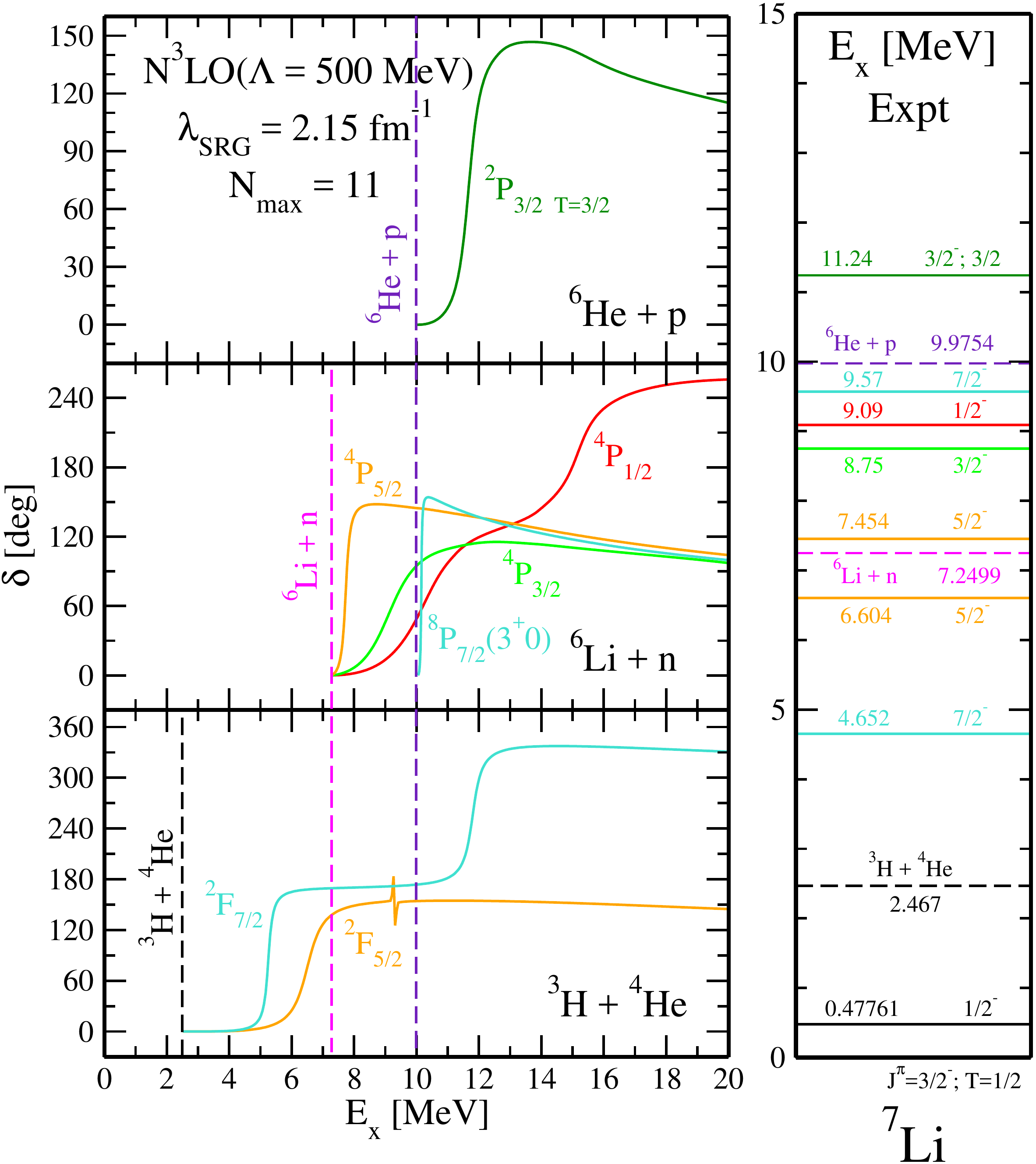}
\caption{\label{spectrum_li}  (Color online) The same as in Fig.~\ref{spectrum_be} for $^7$Li and $^3$H + $^4$He, $^6$Li + n, and $^6$He + p scattering with calculated thresholds in Table~\ref{li_bound}.} 
\end{center}
\end{figure*}

\begin{figure}[t]
\begin{center}
\includegraphics[scale=0.4]{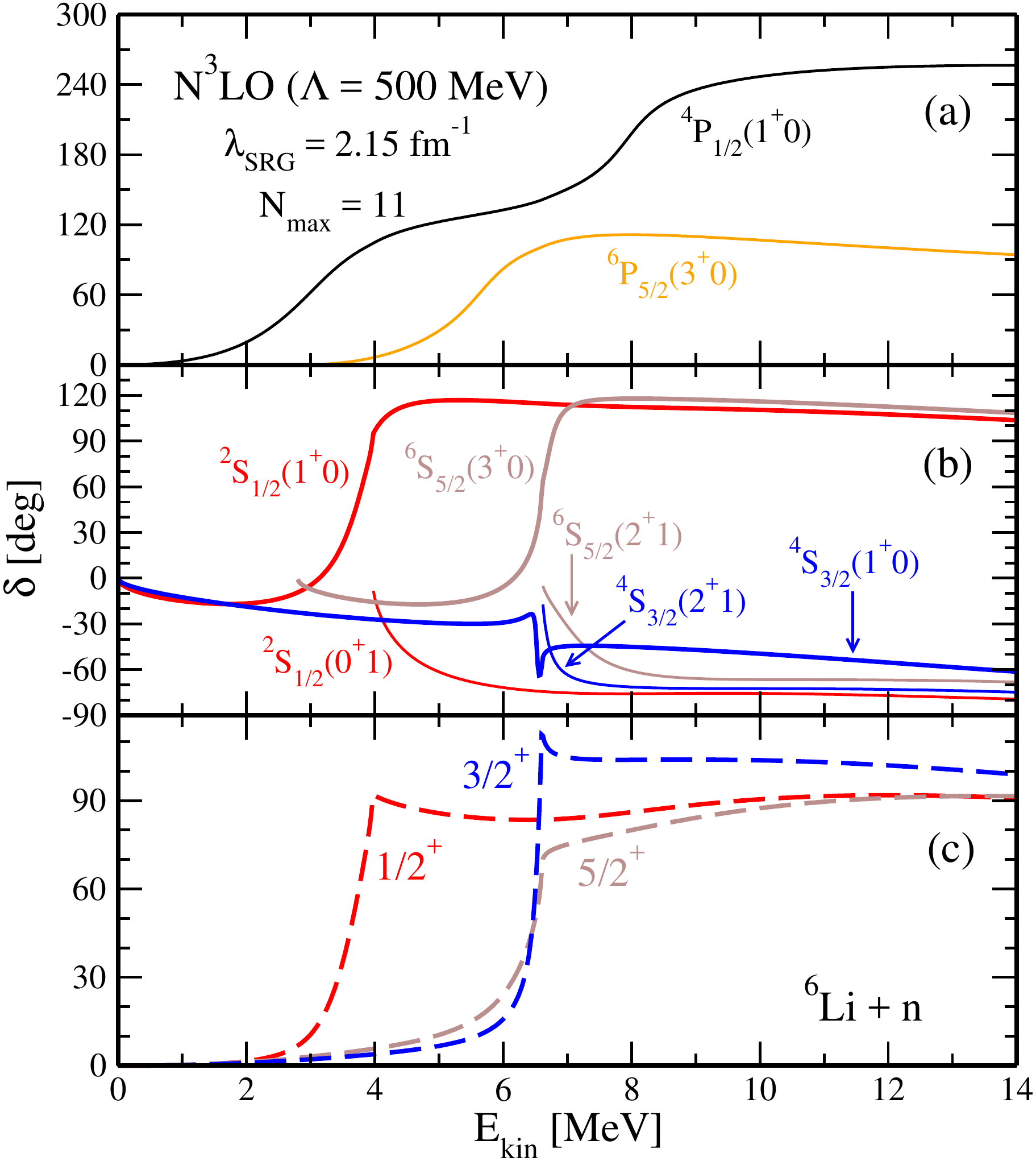}
\caption{\label{li_new_states1}  (Color online) The same as in Fig.~\ref{be_new_states} for $^7$Li from the NCSMC calculation of the $^6$Li+n scattering.}
\end{center}
\end{figure}

\begin{figure}[t]
\begin{center}
\includegraphics[scale=0.4]{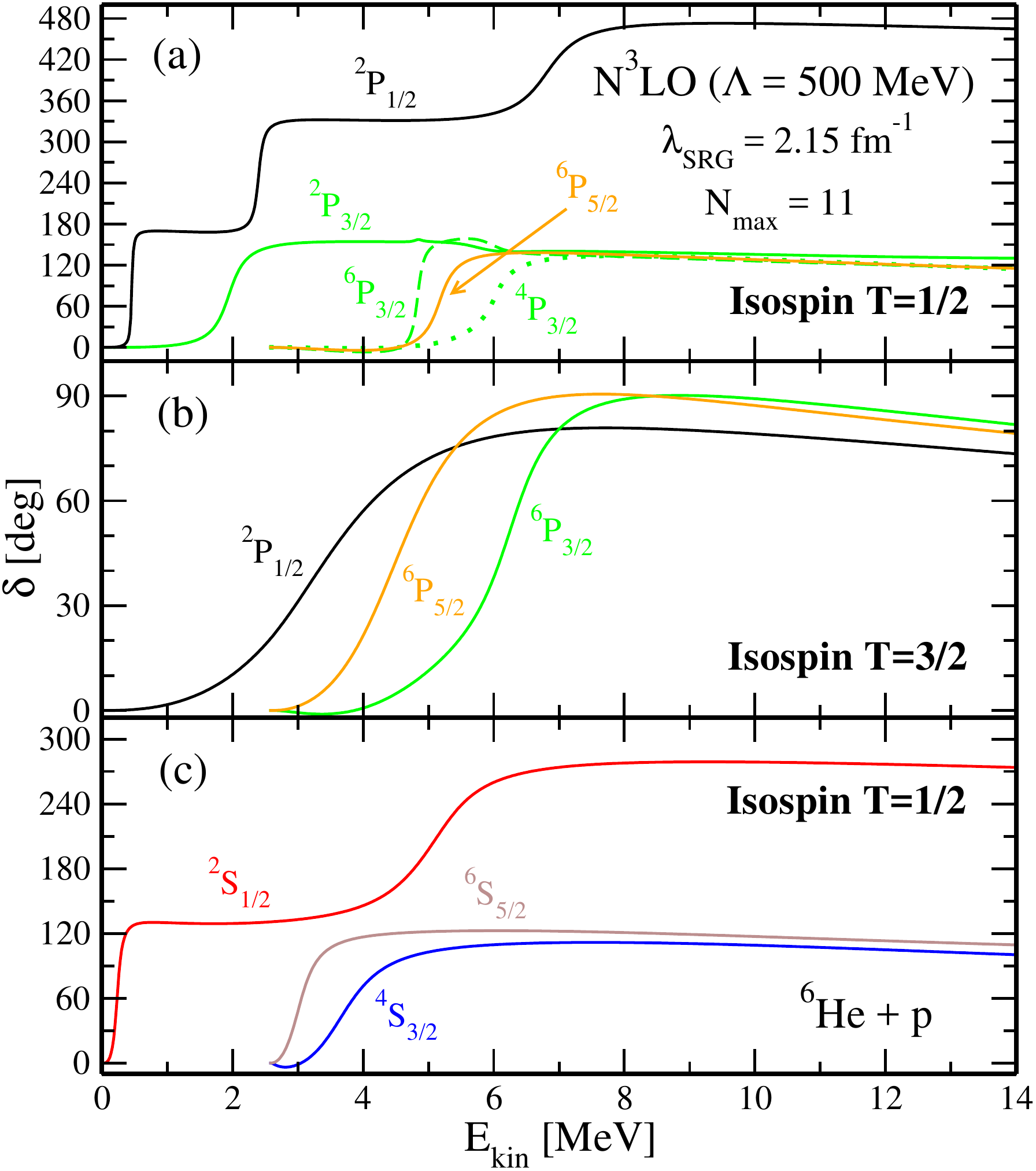}
\caption{\label{li_new_states2}  (Color online) Predictions of new negative- and positive-parity resonant states in the spectrum of $^7$Li obtained from the NCSMC calculation
of the $^6$He + p scattering. The phase shifts are displayed as functions of kinetic energy in the centre-of-mass. All calculations were performed at $N_{\mathrm{max}} = 11$ in the HO
expansion and using the SRG-evolved NN N$^3$LO chiral interaction~\cite{Entem2003} at the resolution scale of $\lambda_{\mathrm{SRG}} = 2.15$ fm$^{-1}$.}
\end{center}
\end{figure}

\begin{table}[t]
\begin{center}
\begin{ruledtabular}
\begin{tabular}{cccccc}
$^3$H + $^4$He & NCSMC & & Exp. & & \\
\hline
$J^{\pi}$ & $E_r$ & $\Gamma$ & $E_r$ & $E_x$ & $\Gamma$ \\
$7/2_1^-$ & 2.79 & 0.214 & 2.18 & 4.652  & 0.069 \\
$5/2_1^-$ & 4.04 & 0.785 & 4.14 & 6.604 & 0.918  \\
$7/2_2^-$ & 9.33 & 0.435 & 7.10 & 9.57 & 0.437 \\
\hline
\hline
$^6$Li + n & NCSMC & & Exp. & & \\
\hline
$J^{\pi}$ & $E_r$ & $\Gamma$ & $E_r$ & $E_x$ & $\Gamma$ \\
$5/2_2^-$ & 0.48 & 0.21 & 0.20 & 7.454 & 0.080 \\
$3/2_2^-$ & 1.83 & 1.70 & 1.50 & 8.75 & 4.712 \\
$1/2_2^-$ & 2.60 & 2.44 & 1.84 & 9.09 & 2.752 \\
$7/2_2^-$ & 2.91 & 0.039 & 2.32 & 9.57 & 0.437 \\
\hline
\hline
$^6$He + p & NCSMC & & Exp. & & \\
\hline
$J^{\pi}$ & $E_r$ & $\Gamma$ & $E_r$ & $E_x$ & $\Gamma$ \\
$3/2_3^-$ & 1.74 & 0.63 & 1.26 & 11.24 & 0.26 \\
\end{tabular}
\end{ruledtabular}
\caption{Energies ($E_r$) and widths ($\Gamma$) in MeV of the resonant states of $^7$Li computed with the NCSMC and compared with the existing experimental
data~\cite{TILLEY20023}. The resonance energies are given with respect the threshold of the corresponding mass partition. All the states have isospin $T=1/2$ except the
$3/2_3^-$ state in $^6$He + p which has $T=3/2$.}  
\label{li_resonances}
\end{center}
\end{table}

\begin{table}[t]
\begin{center}
\begin{ruledtabular}
\begin{tabular}{ccc}
\hspace{0.6cm} $^6$Li + n & NCSMC & \hspace{0.6cm} \\
\hline
\hspace{0.6cm} $J^{\pi} T$ & $E_r$ & $\Gamma$ \hspace{0.6cm} \\
\hspace{0.6cm} $5/2^- \, 1/2$ & 4.79 & 7.73 \hspace{0.6cm} \\
\hspace{0.6cm} $1/2^- \, 1/2$ & 7.71 & 6.13 \hspace{0.6cm} \\
\hspace{0.6cm} $1/2^+ \, 1/2$ & 3.78 & 1.11 \hspace{0.6cm} \\
\hline
\hline
\hspace{0.6cm} $^6$He + p & NCSMC & \hspace{0.6cm} \\
\hline
\hspace{0.6cm} $J^{\pi} T$ & $E_r$ & $\Gamma$ \hspace{0.6cm} \\
\hspace{0.6cm} $1/2^+ \, 1/2$ & 0.23 & 0.13 \hspace{0.6cm} \\
\hspace{0.6cm} $3/2^- \, 1/2$ & 1.94 & 0.41 \hspace{0.6cm} \\
\hspace{0.6cm} $1/2^- \, 3/2$ & 3.03 & 2.65 \hspace{0.6cm} \\
\hspace{0.6cm} $5/2^- \, 3/2$ & 4.43 & 2.10 \hspace{0.6cm} \\
\hspace{0.6cm} $3/2^- \, 3/2$ & 4.55 & 5.21 \hspace{0.6cm} \\
\end{tabular}
\end{ruledtabular}
\caption{Energies ($E_r$) and widths ($\Gamma$) in MeV of some of the new predicted resonant states of $^7$Li computed within the NCSMC. The resonance energies are given with
respect the threshold of the $^6$Li + n and $^6$He + p mass partitions.}  
\label{li_new_resonances}
\end{center}
\end{table}

We present now the results for $^7$Li that we obtained analyzing the $^3$H + $^4$He, $^6$Li + n, and $^6$He + p reactions. In this case there is one more reaction than
in the $^7$Be case and also here these are the all possible binary mass partitions involved in the formation of the $^7$Li system. Also in this case, the three processes
were studied separately.

In Tab.~\ref{li_bound} we summarize the values of the relative and total energies of the two $3/2_1^-$ and $1/2_1^-$ bound states. Since the
$^3$H + $^4$He reaction has the lowest energy threshold, the experimental values are well reproduced and the difference between the total energies is less than 1 MeV, similarly as for $^3$He + $^4$He mass partition in $^7$Be.
The agreement is also good in the other two cases even if the differences are a bit larger.

In Tab.~\ref{li_moments} we display the properties of the ground state obtained from the study of the $^3$H + $^4$He reaction with the NCSM and NCSMC.
The theoretical results for $r_{\mathrm{ch}}$, $Q$, and $\mu$ are compared with the experimental values,
and the theoretical predictions of B(M1; $3/2_1^- \rightarrow 1/2_1^-$) and B(E2; $3/2_1^- \rightarrow 1/2_1^-$) are reported. As in the $^7$Be case, the  $r_{\mathrm{ch}}$, $Q$, and B(E2) values increase substantially when NCSMC is applied due to the proper physical tail of the NCSMC wave functions. We also present the NCSMC ANCs for the ground state and the $1/2^-_1$ state that compare quite well with the values extracted from experimental data. The cluster form factors for the two states resamble closely the $^7$Be ones shown in Fig.~\ref{be_clusterff}. We therefore do not show them.

In Fig.~\ref{spectrum_li} we show the results for the phase shifts. The figure is basically organized as Fig.~\ref{spectrum_be}, here the difference is that we have
three mass partitions instead of two. On the left-hand side there are the three panels displaying the phase shifts of the corresponding process, while on the right-hand side there is the
experimental spectrum. Again, the solid lines represent the energy levels while the dashed lines show the reaction thresholds. The theoretical phase shifts in three panels on the left
are adjusted to the experimental thresholds displayed with dashed lines. As in the previous case, also for $^7$Li our method is able to reproduce all the energy
levels in the correct order. Two differences must be addressed with respect the previous case. The first difference concerns the $T=3/2$ state at $11.24$ MeV, that we now discuss and demonstrate that it is theoretically
well reproduced in particular in the $^6$He + p scattering. The second difference concerns the two $5/2^-$ states at the energies of $6.604$ and $7.454$ MeV, respectively.
In this case the threshold of the $^6$Li + n reaction is exactly in between these two states and thus only the $5/2_2^-$ state appears in the spectrum of this mass partition.
The $5/2_1^-$ resonance is shown in the spectrum of the $^3$H + $^4$He scattering, and once again we find a situation similar to the previous one. The experimental cross section
for $^3$H + $^4$He has a peak in correspondence of the $5/2_1^-$ state, while for $^6$Li + n the peak is found at the energy of the $5/2_2^-$ state. This experimental observation
is reproduced by our calculation and the very small contribution to the $5/2_2^-$ state from the $^3$H + $^4$He process can be seen in its spectrum at the excitation energy
of $\sim 9.5$ MeV. The last comment concerns the $^8\mathrm{P}_{7/2}$ phase shifts in $^6$Li + n scattering, that is built on the $3^+$ state of $^6$Li and in the figure seems
to appear at the threshold of $^6$He + p. This is purely accidental and simply due to the shift of the theoretical results to the experimental thresholds.

In Tab.~\ref{li_resonances} we report the energies and widths of the resonant states of $^7$Li computed with the NCSMC and compared to the experimental values.
For $^3$H + $^4$He reaction we do not include the $5/2_2^-$ state. The general agreement with the experimental values is good, and the resonant centroids are basically
reproduced by our calculations. In this case, the energy of the $7/2_2^-$ state is better reproduced by the $^6$Li + n calculation, even though its contribution to the width
is very small.

Also for $^7$Li our calculations predict new resonant states that are shown in Fig.~\ref{li_new_states1} for $^6$Li + n, and in Fig.~\ref{li_new_states2} for $^6$He + p, respectively. The resonance energies and widths of some of these predicted resonances are summarized in Tab.~\ref{li_new_resonances}. As for $^7$Be, all currently experimentally known $^7$Li states have negative parity, while our NCSMC calculations predict new resonances of both parities. Panel (a) of ~\ref{li_new_states1} shows the $^4\mathrm{P}_{1/2}$ and $^6\mathrm{P}_{5/2}$ phase shifts built on the $^6$Li ground state and the $3^+ 0$ excited state, respectively. We note that the first resonance in $^4\mathrm{P}_{1/2}$ partial wave corresponds to the experimentally known state at 9.02 MeV and the phase shift is also included in Fig.~\ref{spectrum_li}. The narrow resonances in the $^2\mathrm{P}_{1/2}$ partial wave built on the $^6$He ground state presented in Fig.~\ref{li_new_states2} will mix with the much broader $^4\mathrm{P}_{1/2} (1^+ 0)$ partial waves in a calculation that couples different mass partitions as well as in experiment. Consequently, their widths obtained in the present calculations are unrealistically small and we do not include these resonances in Tab. ~\ref{li_new_resonances}. We note that for $T{=}3/2$, we predict a new $1/2^-$ resonance built on the $^6$He ground state as well as a $3/2^-$ and $5/2^-$ resonances built in the $^6$He $2^+ 1$ state, see panel (b) of Fig.~\ref{li_new_states2}. As the latter state is unbound, our predictions for the $3/2^-$ and $5/2^-$ resonances are less robust than that of the $1/2^-$ resonance. The $S$-wave phase shifts built on the four $^6$Li and two $^6$He states included in our calculations are plotted in panel (b) of Fig.~\ref{li_new_states1} and panel (c) of Fig.~\ref{li_new_states2}, respectively. Contrary to the situation in $^7$Be (Fig.~\ref{be_new_states}), the find resonant behavior in the $^6$Li+n scattering in the partial waves built on the $T{=}0$ $1^+$ and $3^+$ $^6$Li states with the resonances appearing below the $T{=}1$ $0^+$ and $2^+$ states thresholds. In the $^6$He+p scattering, the $1/2^+$ resonace appear just above the threshold, also below the $^6$Li $0^+ 1$ state not coupled in the present calculations. Consequently, the prediction of the positive-parity $1/2^+ S$-wave resonance in $^6$He+p appears robust. Still, we have to keep in mind that this state is in the three-body continuum ($^4$He+d+n) that is not included in our calculations and, of course, this can affect its properties.

\subsubsection*{The $^6\mathrm{He} (p, \gamma) ^7\mathrm{Li}$ radiative capture reaction}

The sharp resonance near the threshold of the $^6$He + p reaction suggests a resonant $S$ factor for the $^6\mathrm{He} (p, \gamma) ^7\mathrm{Li}$ radiative-capture reaction.
Indeed, our calculated $S$ factor predicts a very pronounced and sharp peak just above the threshold. Its possible implications for astrophysics, if any, remain to be investigated. As pointed out in the previous subsection, the three-body continuum not included in our calculations could affect this resonance and the $S$ factor.
Since we do not include the coupling between the different mass partitions, the magnitude of our calculated S factor is unrealistically large. Consequently, wedo not present the calculated S factor until the coupling of the $^7$Li mass partitions is implemented in our formalism. Experimental investigation of the $^6\mathrm{He} (p, \gamma) ^7\mathrm{Li}$ radiative capture has been performed only at energies well above the threshold, e.g., at $E_{\rm ^6He}{=}40$ MeV/A~\cite{PhysRevLett.87.042501}. 

\section{Conclusions}
\label{sec_conclusions}

In Ref.~\cite{DOHETERALY2016430} the $^7$Be and $^7$Li systems have been studied within the NCSMC approach investigating the $^3$He + $^4$He and the $^3$H + $^4$He
reactions. In the present work we extended the work of Ref.~\cite{DOHETERALY2016430} and studied these systems in a wider energy range considering all the
binary-mass partitions. In addition to the two previous ones, here we also studied the $^6$Li + p, $^6$Li + n, and $^6$He + p reactions and investigated $^7$Be and $^7$Li bound states as well as resonances and scattering states. Our results provide a very good description of the experimental energy spectrum of both systems.
Not only the bound state energies but also the resonant states are nicely reproduced in the correct order. The widths of the known resonances are also well reproduced. Besides these known states we found several new resonances of both parities, some of them built on the ground state of $^6$Li and $^6$He.
Finally, we also
investigated the $^6\mathrm{Li} (p , \gamma) ^7\mathrm{Be}$ and the $^6\mathrm{He} (p, \gamma) ^7\mathrm{Li}$ radiative-capture processes.
Contrary to the Lanzhou experiment, we did not find any resonance in the $S$-wave near the $^6$Li + p threshold.
Our predicted $^6\mathrm{Li} (p , \gamma) ^7\mathrm{Be}$ $S$ factor is non-resonant and smooth at low energies. On the other hand, we predict a pronounced $S$-wave resonance
in $^7$Li near the $^6$He + p threshold that results in a sharp peak in our predicted $S$ factor for the $^6\mathrm{He} (p, \gamma) ^7\mathrm{Li}$ reaction.
Its possible implications for astrophysics, if any, remain to be investigated. It also must be noted that this state is already in three-body continuum that we did not include in our
calculations and can thus affect our results.

The presented calculations can be improved in three ways. First, the coupling between the different mass partitions needs to be introduced, which would then allow to study transfer reactions such as $^6\mathrm{Li}(\mathrm{n},$$^3\mathrm{H}){}^4\mathrm{He}$. Second, the chiral  three-nucleon interaction shoud be included. Third, the three-body continuum should be considered. The work on the first two points is in progress. The third one is the most challenging and will require more technical development.

\acknowledgments

This work was supported by the NSERC Grant No. SAPIN-2016-00033 and by the U.S. Department of Energy, Office of Science, Office of Nuclear Physics, under Work Proposals
No. SCW1158 and SCW0498. TRIUMF receives federal funding via a contribution agreement with the National Research Council of Canada. This work was prepared in part by LLNL
under Contract No. DE-AC52-07NA27344. Computing support came from an INCITE Award on the Titan supercomputer of the Oak Ridge Leadership Computing Facility (OLCF) at
ORNL, from Westgrid and Compute Canada, and from the LLNL institutional Computing Grand Challenge Program.

\newpage

\end{document}